\documentclass[journal,draftclsnofoot,onecolumn,12pt]{IEEEtran}
\usepackage{amsthm,amssymb,graphicx,multirow,amsmath,color,amsfonts}%,ulem}
\usepackage[update,prepend]{epstopdf}
\usepackage[noadjust]{cite}
\usepackage[latin1]{inputenc}
\usepackage{tikz}
\usetikzlibrary{angles, quotes}
\usepackage{bbm} % for \mathbbm{1}
\usepackage{pdfpages}
\usepackage{tabulary}
\usepackage{multirow}
\usepackage{comment}
\usetikzlibrary{calc}
\usepackage{ amssymb }
\usepackage{subfig}
\usepackage{hhline}
\include{notation}
\allowdisplaybreaks % Allows breaking of eqnarray over multiple pages (avoids unnecessary blanks in the document before eqnarray)
\usepackage{gensymb}

\begin{document}
%\pagenumbering{gobble}
\graphicspath{{./Figures/}}

\title{{\huge On the Topological Aspects of UAV-Assisted Post-Disaster Wireless Communication Networks}}
\vspace{1cm}
\author{
Maurilio Matracia, {\em Student IEEE}, Mustafa A. Kishk, {\em Member IEEE}, and Mohamed-Slim Alouini, {\em Fellow IEEE} 
\thanks{The authors are with Computer, Electrical and Mathematical Science and Engineering Division, King Abdullah University of Science and Technology (KAUST), Thuwal 23955-6900, Saudi Arabia (email: $\{$maurilio.matracia; mustafa.kishk; slim.alouini$\}$@kaust.edu.sa). % remove the date for conference drafts
}}

\maketitle
%\begin{comment}
\vspace{0cm}\begin{abstract}
In the context of sixth-generation (6G) networks, emergency management systems (EMSs) based on wireless communications have recently gained increasing interest.
Hereby, fundamentals and open problems of post-disaster communications are discussed, especially  focusing on their topological aspects.
The motivation behind this choice is due to the fact that, whenever a natural or a man-made disaster occurs, there is a high chance that the terrestrial communication infrastructure is compromised, and therefore alternative networks need to be deployed efficiently in order to enable the majority of the civilians and the first responders (FRs) to communicate.
In this paper, we first provide a brief review of existing aerial ad-hoc networks for post-disaster communications.
Next, we shed light on some new aspects of this problem, which are related to the topology of the network supporting the impacted area.
Finally, with the aid of selected simulation results, we show how the cellular infrastructure requirements for a disaster-struck region significantly depend on its location and its extension.
\end{abstract}
\vspace{2cm}
%\end{comment}

\vspace{3mm}

\section{Introduction} \label{sec:Intro}
A disaster is an event that causes destruction and distress over a large territory. Worldwide, natural as well as man-made disasters are responsible for massive and often unpredictable losses. 
Especially considering that climate change implies more frequent cataclysms, an efficient network-assisted EMS is needed in order to predict and identify the incidences of natural calamities, easing to respond on a timely manner.
Note also that when exceeding 72 hours the chance of survival of the trapped victims rapidly decreases, and hence data dissemination and cellular coverage are indispensable in these types of emergency situations.   \par
Therefore, disaster management (that is, a discipline based on preparation, response, and relief in case of catastrophic situations) procedures have gained attention in the last decade.
All the activities done in disaster management procedures are usually categorized as pre-disaster or post-disaster ones. 
The former essentially consist of prevention, detection, mitigation, and preparedness, and they aim to limit the damage due to a natural adversity by minimizing hazards, for example by informing people about the incoming danger.
On the other hand, post-disaster activities essentially regard damage estimation and repair, recovery, and relocation, and are supposed to last much longer than the former. \par
Our main objective is to study post-disaster networks assisted by aerial base stations (ABSs) from a novel perspective that focuses on their topological aspects.
For example, the authors in \cite{Adams2011} surveyed the use of UAVs in post-disaster scenarios for imagery collection without considering their use as ABSs nor the influence of the topology of the suffered region.
Despite many problems (e.g., trajectory optimization, transceiver design, and intercell-interference-coordination) have been addressed in \cite{Zeng_sky19, Zhao19, Merwaday16}, some important aspects such as the extension and the location of the disaster area have been omitted. \par
The main contributions of this paper are therefore: \\
$\bullet$ Disaster assessment is discussed from a novel topological perspective; \\  
$\bullet$ Different fleets of UAVs are quantitatively compared in terms of ergodic capacity enhancement in a wide range of typical disaster scenarios; \\
$\bullet$ Open problems and conceptual solutions are proposed for intra-region and extra-region emergency communications. \par
In the rest of this paper, we first provide a brief survey on existing solutions for post-disaster aerial communications.
Next, we focus on the topology of typical post-disaster communication networks. 
In particular, we build realistic simulation setups to obtain various insightful results.
Before drawing our conclusions, we also discuss open problems that should be carefully addressed in future to optimize the performances of post-disaster communication networks. 

\vspace{3mm}

\section{Brief Survey on Aerial Backup Networks} \label{sec:Literature}
Whenever a natural or a man-made disaster occurs, the main (terrestrial) communication infrastructure is usually overloaded and subjected to partial or total failure. 
Therefore, it is vital to arrange a reliable backup network in order to ensure sufficient coverage. \par
\begin{figure}
\centering
\includegraphics[width=0.7\columnwidth]{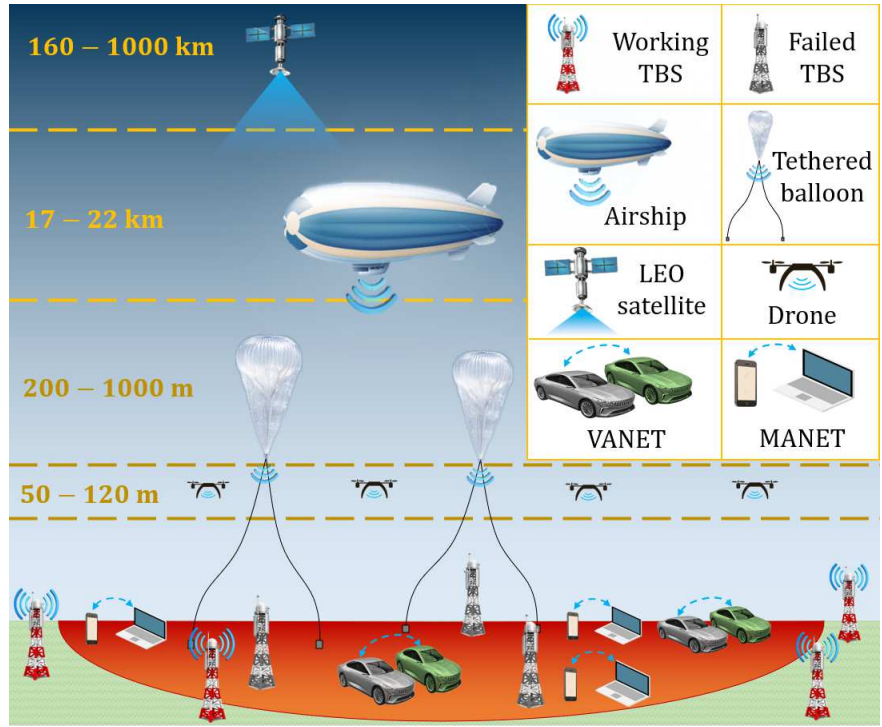}
\caption {Integrated communication systems including non-aerial typical ad-hoc paradigms such as LEO satellites, vehicular ad-hoc networks (VANETs), and mobile ad-hoc networks (MANETs). 
The red area represents the disaster-struck zone.}
\label{fig:Architecture}
\end{figure}

We are moving towards a data-centric (rather than voice-centric) world, hence there is a strong interest in designing cellular-based communications for public safety. 
In fact, some land mobile radio systems take advantage of the latest generations of mobile communications to furnish an adequate emergency broadband service \cite{Zeng2016}. 
However, the costliness of the terrestrial base stations (TBSs) still represents an important limitation, therefore research work has been mainly directed towards studying various alternative networks that might also reduce the overall cost of cellular coverage (some of them are illustrated in Fig. \ref{fig:Architecture}). \par
Capable of facilitating fundamental requirements such as enhanced mobile broadband (eMBB), ultra reliable low latency communications (URLLC), and massive machine type communications (mMTC), unmanned aerial vehicles (UAVs\footnote{Although the terms drone and UAV are used interchangeably in several works, the latter should actually refer to any flying vehicle with no pilot, including gliders, balloons, and airships, for instance.})
can also serve other emergency services \cite{Zeng_sky19, Saad20}. 
However, this technology has not been fully standardized yet and will probably become mature with the advent of 6G \cite{chowdhury20206g}.
Indeed, the lack of ubiquitous low-Earth orbit (LEO) satellite constellations or uptilted TBSs' antennas for providing aerial backhaul, as well as insufficient data rates for ensuring a safe coordination of large fleets, are technical limitations that still need to be overcome while reducing the costs of deployment.
\par
The main paradigms of UAV networks will be discussed in this section.
Manned vehicles such as helicopters and airplanes have been considered for emergency communications as well, but they are now less favorable due to the requirement of a pilot and the concurring improvements of other unmanned technologies.

\subsection{Low-Altitude Networks}
Low-altitude platforms (LAPs) are typically operating in the troposphere layer, hence their altitude does not exceed ten kilometers.
These platforms, however, are not suitable for catastrophes caused by storms or hurricanes because strong wind and rain reduce the payload capacity and the reliability of communications, respectively.
This subsection introduces the main types of LAPs. 

\vspace{2mm}
\subsubsection{Drones} Despite drawbacks such as limited achievable payload and flight time as well as high risk of collisions, drones are actually representing the most promising technology for emergency communications due to their advantages in terms of low cost, high mobility, and  the fact that they do not require any pilot, to name a few. \par
Specific designs for drone stations intended to operate in emergency scenarios have been proposed in \cite{Erdelj2017}, \cite{Merwaday16}.
Authors in \cite{selim2018post} suggested a cooperative fleet of drones made by a grid of ABSs supported by a tethered drone for backhaul and two powering drones for extending their flight time.
As remarked in \cite{Erdelj2017}, image processing and information exchange (e.g., position, flight trajectory, speed) between UAVs are needed to avoid collisions, especially when maneuverability is limited. 
This mostly happens in case of harsh meteorological conditions, for which safe UAV landing procedures would be needed.

\vspace{2mm}
\subsubsection{Tethered Drones} When tethering a drone, higher autonomy and signal strength are achieved at the price of limiting its mobility and relocation flexibility.
The optical fiber connection integrated in the tether ensures a better backhaul link and the power supply enhances the flight time from an hour to even more than a month.
Note also that having a limited mobility might be an advantage in case of harsh meteorologic conditions, since the risk of having a collision is drastically reduced \cite{Kishk19}.

\vspace{2mm}
%\begin{comment}
\subsubsection{Low-Altitude Balloons}
Interesting research has been carried on to support the integration of light-fidelity (LiFi) systems on low-altitude balloons \cite{Surampudi2018a} to conduct search and rescue (SAR) operations, showing that LiFi-equipped balloons can represent a cheap and more versatile alternative to radio-frequency (RF) systems  (more versatile because LiFi technologies can also work in case of floods or in presence of combustible gas leakages). 
Moreover, this solution may also provide sufficient illumination to rescue teams and victims.
%\end{comment}

\vspace{2mm}
\subsubsection{Tethered Balloons} There are several reasons for equipping balloons with a tethered link, such as the better balloon controllability, the more stable power supply, and the more reliable data link through the tether.
Tethered balloons can be rapidly deployed at considerable altitudes (up to a kilometer) above the sea level, thus ensuring a considerable coverage area. 
The effectiveness of such ABSs for disaster mitigation and response has been proved in \cite{Alsamhi2018}. 

\subsection{High-Altitude Networks}
High altitude platforms (HAPs), also referred to by "high altitude airships", "stratospheric platforms", or "atmospheric satellites", present various advantages such as wide coverage areas (i.e., tens of km$^2$), favorable channel characteristics, and ease of deployment. 
Usually, HAPs are deployed at altitudes of 17-22$\,$km. At these altitudes, the energy required for levitating is strongly reduced because of the low turbulence and wind currents.
In addition, due to a better exposure to the Sun, solar panels and batteries can extend their flight times up to several months.\par
Compared to satellites, HAPs have the advantages of lower latency, persistence (i.e., the ability to hover over the same area for a long period of time), better image resolution (since they are closer to the ground), and the ability to go back to their base for maintenance or payload reconfiguration \cite{Grace2011}.
Finally, HAPs are generally faster to deploy than both TBSs and satellites, since they just require few hours to start operating.
The most relevant types of HAPs are balloons, airships, and gliders.

\vspace{2mm}
\subsubsection{Balloons} On the contrary of tethered balloons, the conventional ones are usually working as HAPs, since the absence of cables allows them to freely reach higher altitudes (approximately 20$\,$km).
Because of this, their coverage radius can exceed 70$\,$km. \par
Implementing this technology is still very expensive, indeed "Project Loon" from Alphabet (i.e., Google's parent company), which aimed to develop a network of high altitude super-pressured solar balloons to furnish Internet access in remote areas, has been shut down in early 2021.

\vspace{2mm}
\subsubsection{Airships} Interesting projects have been initialized to develop airships that address all the requirements for emergency scenarios. 
One of the most promising innovations, Skyship, has been proposed by KT Corp.
The project aims to realize an innovative platform for safety management. 
The platform carries a set of drones and car robots equipped with cameras, which are released from the platform to investigate closely the affected area and to establish video calls between the civilians and the first responders.
Skyship would also be connected by means of 5G or LTE technologies to a station on the ground for command, control and communication.

\vspace{2mm}
\subsubsection{Gliders}
Without using fuel or helium, gliders need batteries and a vast surface of PV modules in order to stay aloft for a long time.
Consequently, their design becomes extremely complex compared to other HAPs, since the wide fixed wings will be exposed to intense mechanical stresses at very low temperatures.
Finally, the use of fixed wings limits the persistence of the aircraft, although it allows reaching the disaster area more rapidly.
Despite all the aforementioned challenges, the idea of using a glider-mounted base stations (BSs) was realized by NASA's Pathfinder-Plus in 2002 and now companies such as Airbus, UAVOS, and Softbank's HAPSMobile are improving their respective gliders (Zephyr, HAPS, and Sunglider) in order to commercialize them soon.

\vspace{3mm}

\section{Topological Aspects for Disaster Assessment} \label{sec:DisasterAssessment}
Whenever a disaster strikes, the very first counteraction to make is definitely to estimate the damage it caused.
Keeping in mind the brief overview provided in Sec. \ref{sec:Literature}, hereby we introduce the main topological aspects to consider for choosing the best type of aerial platforms for post-disaster scenarios.
Note that UAVs can also play a crucial role by means of the remote sensing instrumentation they can be equipped with.
Indeed, camera-equipped UAVs can determine many features of the disaster area (e.g., hazard maps, dense surface models, detailed building renderings, comprehensive elevation models).
Most recent technologies also enable automated map creation in order to rapidly assess the damages inside the suffered region \cite{Adams2011}.
The communication techniques for disaster assessment, however, are out of the scope of this paper.

\subsection{Extension of the Disaster Area} \label{subsec:extension}
Estimating the size of the disaster area is vital to succeed in emergency missions.
Note that the most diffused technologies for cellular communications are susceptible to disasters, since their infrastructure can take months to recover after a serious failure. \par
As stated in the Sec.\ref{sec:Literature}, there are various types of aerial platforms to deploy in disaster-struck regions in order to enhance cellular coverage, each one with its own advantages and disadvantages. 
The knowledge of the extension of the affected region is vital to determine which type of aerial platforms best suits the affected area.
In addition, the size of such area determines whether an action is required to be taken or, instead, the surrounding cellular infrastructure can provide sufficient coverage.

\subsection{Geographical Location}
\label{subsec:location}
In order to plan the correct aid for searching and rescuing all the victims trapped inside the disaster region, also detailed information about the geographical location of the site is needed.
For instance, a confined disaster area located in the periphery of a small town could still count on the surrounding ground BSs, whereas if a tragedy happens in an isolated village it will be definitely preferable to just send a fleet of drones to support it. 
Furthermore, when streets and roads are interrupted by debris or even destroyed, ABSs are preferred to traditional ground response systems because they can speed up the process of identification of victims, avoiding limited communication services relying on large latency and low bandwidth satellite communications. 

\begin{figure*}
\centering
\makebox[\textwidth][c]{\includegraphics[width=1\columnwidth, trim={0cm 0cm 0cm 0cm},clip]{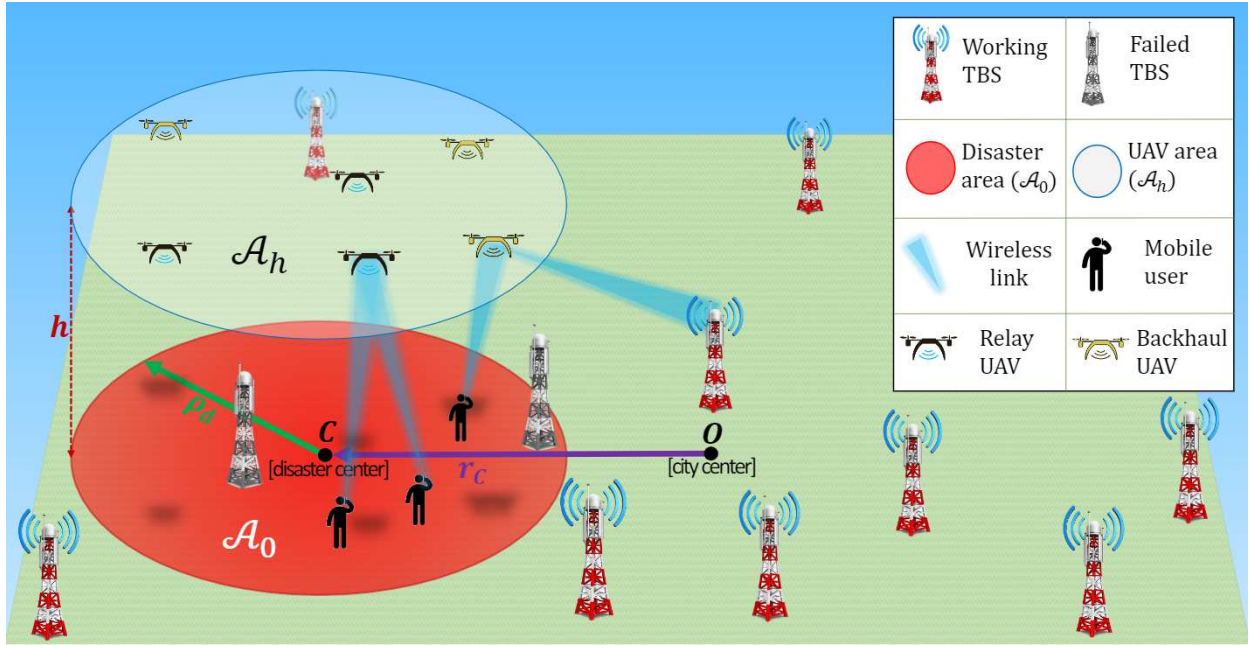}}
\caption {Proposed system setup.
The TBSs inside the suffered region (assumed circular with radius $\rho_d$) fail, and thus a fleet of UAVs is deployed into the same region either to provide cellular coverage or for assessment. }
\label{fig:intra-extra}
\end{figure*}

\subsection{Load Distribution}
The spatial distribution of the users in the affected region highly affects the selected type of backup wireless infrastructure that should be deployed after a disaster.
For instance, if the users are sufficiently clustered around gathering points, deploying one single aerial platform at relatively high altitude above each cluster might conveniently imply having negligible interferences.
If users are very sparse, instead, the best option might be to deploy a large number of drones in order to locally assist both SAR operations and cellular service enhancement.

\subsection{Network Resilience and State of the Infrastructure} \label{subsec:resilience}
The level of resiliency of a network can be expressed by the quality of resilience (QoR) metric, a factor between $0$ and $1$ that directly quantifies the ability to react in failure instances without being perceived by users \cite{deepak2019overview}.
Evidently, the state of the existing infrastructure must be known prior sending ABSs to support the users.
For example, some TBSs inside the disaster-struck zone could still work despite the failure of the other ones, hence the locations and conditions of the BSs in the suffered region should be taken into consideration when selecting the type, number, and locations of backup network nodes. 
This concept has been highlighted in \cite{Zhao19}, where the framework of UAV-assisted emergency networks in disaster situations depends on the state of the infrastructure.
If there are still some active TBSs, ABSs may cooperate with them in order to serve the terrestrial users.

\vspace{3mm}

\section{Network Performances in  Post-Disaster Scenarios}
This section aims to introduce the topological aspects defined in Sec. \ref{sec:DisasterAssessment} in a realistic model simulating a disaster-struck terrestrial network assisted by specific fleets of ABSs.

\subsection{System Setup} 
We model the locations of the nodes as point processes (PPs), which is a common assumption in literature \cite{Zeng_sky19,hmamouche2021new}.
In relation to Sec. \ref{subsec:resilience}, we assume the worst case, that is a network with null QoR.
To compensate the complete breakdown of the TBSs within the disaster area, represented in Fig. \ref{fig:intra-extra} by a circle $\mathcal{A}_0$ with radius $\rho_d$ and centered at distance $r_c$ from the town center, a fleet of $n_A$ identical ABSs is deployed uniformly at fixed altitude $h$ above the disaster area (i.e., scenario 1 in \cite{Zhao19}).
Each UAV is either in line-of-sight (LoS) or non-line-of-sight (NLoS) condition with respect to the user, independently from the other UAVs.\par
All the transmitted signals are assumed to experience standard power-law path loss propagation model with exponent $\alpha$ depending on the type of transmitter, namely TBS, LoS UAV, or NLoS UAV.
The serving BS is the one that provides the maximum average received power, whereas all the other BSs are assumed to interfere.

\subsection{Simulation Results}
Assuming a dense urban environment, two sets of simulations are proposed: (i) the first one assumes homogeneous Poisson PP (HPPP)-distributed TBSs, and hence, the location of the disaster is irrelevant, 
(ii) the second one considers inhomogeneous Poisson PP (IPPP)-distributed TBSs which density decreases when moving away from the town center, thus the distance of the disaster from the origin (i.e., the town center) is relevant. 
We evaluate the ergodic capacity, referred to as $R$, that can be achieved by means of various fleets of drones, tethered balloons, or HAPs, for various values of $n_A\,$.
The ergodic capacity is the average Shannon capacity given that the signal to interference plus noise ratio (SINR) exceeds a certain threshold $\tau$.
Averaging over $10^5$ iterations of the PPs, $R$ is evaluated in fifty points as function of $\rho_d$ for the HPPP setup and of $r_c$ for the IPPP one.
Indeed, the main scope of these setups is to investigate the influences of the extension and the location of the disaster area respectively introduced in Secs. \ref{subsec:extension} and \ref{subsec:location}. \par
Our main question is the following: what is the best fleet of aerial platforms (among drones, tethered balloons, and HAPs) given $\rho_d$ and $r_c$?
In what follows we show that the answer is not trivial due to the inherent trade-off that exists between providing higher signal quality for ground users and causing higher interference for mobile users served by other BSs.\par
For the proposed simulations, the bandwidth, the SINR threshold, and the noise power spectral density are set equal to 100$\,$MHz, -5 dB, and 10$^{-12}$ W/Hz, respectively.
Furthermore, for TBSs, drones, tethered balloons, and HAPs, the transmit power $p_t$ is equal to 10, 1.585, 10, and 20 [W] while the altitude $h$ is set to 0, 0.1, 0.5, and 17 [km], respectively. 
Finally, the exponent $\alpha$ and the Nakagami-m shape parameter $m$ equal 2 for LoS ABSs and respectively equal 3 and 1 otherwise, whereas the mean additional losses $\eta$ are set to 0.005 for NLoS ABSs and 0.692 otherwise.

\vspace{2mm}
\subsubsection{HPPP-distributed TBSs}

\begin{figure}
\centering
\subfloat[]{\includegraphics[width=0.5\columnwidth, trim={1cm 0cm 1.6cm 0.9cm},clip]{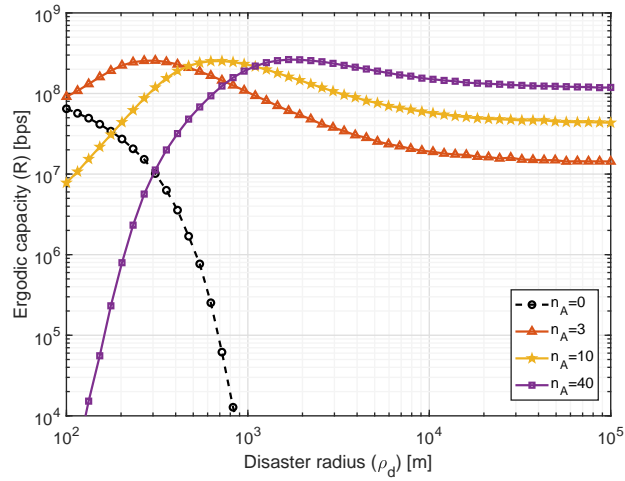}\label{fig:hTBSs.a}} \\
\subfloat[]{\includegraphics[width=0.5\columnwidth, trim={1cm 0cm 1.6cm 0.9cm},clip]{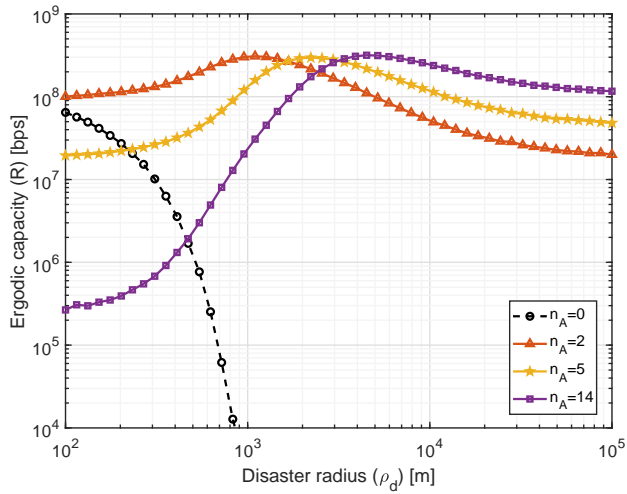}\label{fig:hTBSs.b}}\\
\subfloat[]{\includegraphics[width=0.5\columnwidth, trim={1cm 0cm 1.6cm 0.9cm},clip]{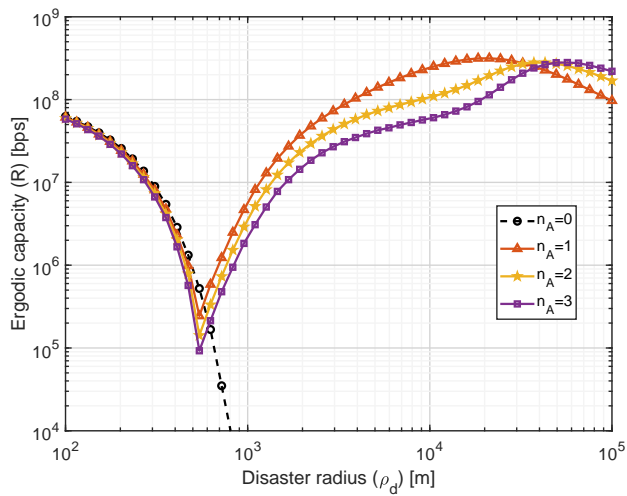}\label{fig:hTBSs.c}}%
\caption {Ergodic capacity as function of the disaster radius in case of HPPP-distributed TBSs assisted by: (a) drones, (b) tethered balloons, and (c) HAPs.}
\label{fig:hTBSs}
\end{figure}

In the first set of simulations, the locations of the TBSs are modeled as a HPPP of density 10$\,$BSs/km$^{-2}$.
The results are illustrated in Fig. \ref{fig:hTBSs} when drones, tethered balloons, or HAPs are uniformly distributed over $\mathcal{A}_h$. 
For the three cases, the black curves are equivalent because they refer to the same system with no UAVs.
On the other hand, different nonzero values of $n_A$ have been chosen for each type of vehicles, since they have completely different coverage areas. \par
Various insights can be drawn from Fig. \ref{fig:hTBSs}, as follows: \\
$\bullet$ When $\rho_d$ exceeds a few hundred meters, the terrestrial infrastructure becomes insufficient, since $R$ monotonically decreases due to the increasing path loss; \\
$\bullet$ In general, the optimal $n_A$ decreases as $h$ and $p_t$ increase; \\
$\bullet$ When deploying LAPs, from Figs. \ref{fig:hTBSs.a} and \ref{fig:hTBSs.b} we can observe that at low values of $\rho_d\,$, deploying just a few ABSs is most favorable.
This is mainly because more numerous fleets generate a stronger interference. 
However, as $\rho_d$ increases, many more ABSs (especially if drones are being deployed) are needed to cover $\mathcal{A}_0\,$; \\
$\bullet$ For values of $\rho_d$ up to 1$\,$km it is discouraged deploying any HAPs, as shown in Fig. \ref{fig:hTBSs.c}. 
Indeed, in this scenario the user has almost no chance of benefiting from HAPs (despite the latters' $p_t$ being twice its TBSs' counterpart) because there is an excessive difference in terms of their respective minimum distances from the user, namely $17\,$km for HAPs as opposed to $\rho_d\,$ for TBSs.
This makes HAPs essentially operate as interferers, thus worsening the performances of the network. \par
On the other hand, when $\rho_d$ exceeds a couple of kilometers the user is much less influenced by the surrounding TBSs, and can now take advantage of the support offered by the available HAP(s).

\begin{figure}
\centering
 {\includegraphics[width=0.55\columnwidth, trim={1cm 0cm 1.6cm 0.9cm},clip]{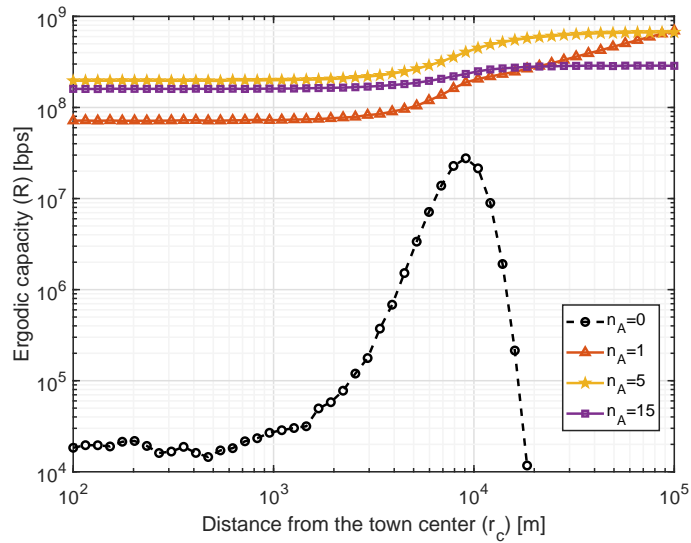}}\\
 \caption{Ergodic capacity as a function of the distance between the disaster and the town center when TBSs are IPPP-distributed and $\rho_d$ equals 500 m. }%
 \label{fig:iTBSs.b}%
\end{figure}

\vspace{2mm}
\subsubsection{IPPP-Distributed TBSs}  
The locations of the TBSs have been generated according to a bidimensional Gaussian distribution with variance 10$\,$km$^2$ centered around the town center, and such that the average number of TBSs within a distance of 100$\,$km from the town center is $1254$.
Assuming $\rho_d$ equal to $500\,$m, Fig. \ref{fig:iTBSs.b} describes $R$ as function of $r_c$ when deploying $n_A$ drone BSs.
The following insights can be extracted:\\
$\bullet$ By looking at the behavior of $R$, we note that the steepest increase always occurs when $r_c$ is approximately three times the standard deviation of TBSs' distribution.
At this value of $r_c$, indeed, the terrestrial interference becomes weak, as evident also from the case with no ABSs (black curve);\\
$\bullet$ As expected, deploying a larger number of drones reduces the dependence on $r_c$, since it reduces the association probability to TBSs as well as the influence of terrestrial component on the aggregate interference;\\
$\bullet$ The optimal $n_A$ never reaches fifteen and actually tends to decrease to one when the disaster is moved far away from the town center.
Indeed, deploying even a single drone is usually preferable when the TBSs are sparse, because it keeps the serving distance short while bringing the aerial interference to zero.\par
In conclusion, it is convenient to deploy a single drone when the suffered region is located very far from the town center and is confined within approximately $1\,$km$^2$.
However, the only solution for much vaster failures would be to increase $n_A$ (as already noted in Fig. \ref{fig:hTBSs}) independently from $r_c$, since a very large homogeneous area surrounds the typical user.

\vspace{3mm}

\section{Intra-Region and Extra-Region Communications} \label{sec:Intra-Extra}
An interesting open problem consists in providing a system architecture that enhances the quality of communication (i) among users inside the affected region, and (ii) with the outer world through the cellular infrastructure.
These two targets may require different technologies and devices. 
For example, it is easy to imagine fire fighters communicating by portable two-way radio devices (i.e. walkie-talkies operating on public safety bands) when operating inside a burning building, but a longer range technology such as cellular would be indispensable for extra-region communications. \par 
According to Fig. \ref{fig:intra-extra}, UAVs can be used as intra-region communication relays for assessment to allow users to share safe routes, shelters locations, and maps, or can be used to establish a backhaul link to provide cellular coverage and connect the users inside the disaster zone to the outer network. 
In the latter case, ABSs' locations depend on the availability of backhaul links, and hence need to be optimized for maximizing networks' performances.\par
Intuitively, backhaul ABSs should be placed in the periphery of the disaster in order to be as close as possible to the surrounding TBSs, but nonetheless further studies are needed to prove this and to investigate the influence of the load distribution and the other topological aspects introduced in Sec. \ref{sec:DisasterAssessment}. 
Another option could also be to make the ABSs fly at different altitudes: the ones needed for backhaul could fly at lower altitudes, compatibly to LoS conditions, in order to reduce the distance to the closest TBS.

\vspace{3mm}

\section{Challenges, Conclusions, and Future Works} \label{sec:Conclusions}
Since no static telecommunication infrastructure can be considered reliable when exposed to harsh calamities, implementing ad-hoc aerial networks in a timely manner is an important goal to target for future 6G networks.
In fact, some crucial technical challenges still need to be solved: 
for example, tethered LAPs and HAPs have limited mobility and often require excessive time for being deployed.
On the other hand, untethered LAPs are limited by a short flight time and struggle in finding reliable backhaul links.
Finally, managing large number of aerial vehicles while avoiding obstacles and optimizing their trajectory is a complex task that still requires further research. \par
In this work, we have discussed the main topological aspects of disaster assessment and network architectures in order to evaluate the ergodic capacity and determine the best aerial fleets to deploy in typical emergency situations.
In future, more complex setups where UAVs' locations and cardinality are optimized depending on the topology of the network should be considered.
Finally, we have shown that a similar approach can be useful for intra- and extra-region communications management, which also deserve further studies.

\bibliographystyle{IEEEtran}
\bibliography{References}

\vspace{1cm} 

\begin{IEEEbiographynophoto}
{Maurilio Matracia}
[S'21] is a Ph.D. student in the communication theory lab (CTL) at King Abdullah University of Science and Technology (KAUST). 
He received his B.Sc. and M.Sc. degrees in Energy and Electrical Engineering from the University of Palermo (UNIPA) in 2017 and 2019, respectively.
His main research interest is stochastic geometry, with special focus on rural and post-disaster communications.
\end{IEEEbiographynophoto}

\vspace{-1cm}

\begin{IEEEbiographynophoto}
{Mustafa A. Kishk} [S'16, M'18] is a postdoctoral research fellow in the
CTL at KAUST. 
He received his B.Sc. and M.Sc. degree from Cairo University in 2013 and 2015, respectively, and his Ph.D. degree from Virginia Tech in 2018.
His current research interests include stochastic geometry, energy harvesting wireless networks, UAV-enabled communication systems, and satellite communications.
\end{IEEEbiographynophoto}

\vspace{-1cm}

\begin{IEEEbiographynophoto}
{Mohamed-Slim Alouini} [S'94, M'98, SM'03, F'09] was born in Tunis, Tunisia.
He received the Ph.D. degree in Electrical Engineering from the California
Institute of Technology (Caltech), Pasadena, CA, USA, in 1998. 
He is now a Distinguished Professor of Electrical Engineering at KAUST.
His current research interests include the modeling, design, and performance analysis of wireless communication systems.
\end{IEEEbiographynophoto}

\end{document}